\documentclass{phb-proc4-auth}

\usepackage{graphicx}
\usepackage{amssymb}

\begin{document}
\begin{frontmatter}


\journal{SCES '04}


\title{Solid state Pomeranchuk effect}


\author[uff]{M. A. Continentino\thanksref{ABC}\corauthref{1}}
\author[uff]{, A. S. Ferreira}
\author[camp]{, P. G. Pagliuso\thanksref{CDE}}
\author[camp]{, C.  Rettori}
\author[US]{, J. L. Sarrao}


\address[uff]{Instituto de F\'{\i}sica, Universidade Federal Fluminense,
Niter\'oi, RJ, 24.210-340, Brazil.}
\address[camp]{Instituto de F\'{\i}sica ``Gleb Wataghin'', UNICAMP,
13083-970, Campinas, Brazil}
\address[US]{Los Alamos National Laboratory, Los Alamos, New Mexico 87545,
USA.}


\thanks[ABC]{Work supported by CNPq-Brasil (PRONEX98/MCT-CNPq-0364.00/00)
and FAPERJ}

\thanks[CDE]{Work supported by CNPq-Brasil (Grant 3076668/03) and
FAPESP}

\corauth[1]{Corresponding Author: Instituto de F\'{\i}sica,
Universidade Federal Fluminense, Niter\'oi, RJ, 24.210-340,
Brazil. Phone: (55-21) 2629-5816 Fax: (55-21) 2629-5887, Email:
mucio@if.uff.br}


\begin{abstract}
Recently we have shown that $YbInCu_{4}$ and related compounds
present a solid state Pomeranchuk effect. These systems have a
first order volume transition where a local moment phase coexists
with a renormalized Fermi liquid in analogy
with $%
^3He$ at its melting curve.  We demonstrate here experimentally
that the solid state Pomeranchuk effect, controlled by a magnetic
field, can be used to produce cooling.
\end{abstract}

\begin{keyword}
Kondo lattice \sep Pomeranchuk effect
\end{keyword}

\end{frontmatter}

\begin{figure}[th]
\begin{center}
\includegraphics[height=4.6cm]{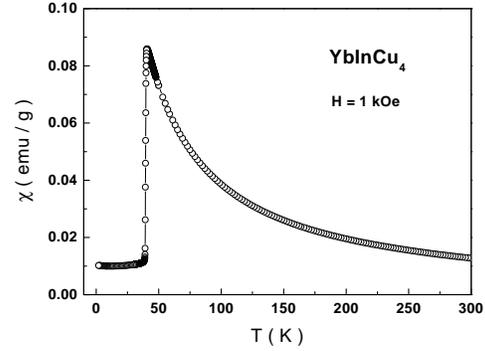}
\end{center}
\caption{ Magnetic susceptibility of $YbInCu_{4}$ as a function of
temperature measured in a field of $1$ kOe. At $T_{V}\approx 42$ K
Fermi liquid and local moment phases coexist. } \label{fig1}
\end{figure}

The system $^{3}He$ along its melting line presents the unusual
feature that the entropy of the liquid is smaller than that of the
solid \cite{betts}. This is the basis of the Pomeranchuk effect
\cite{betts} which has an important application as a cooling
mechanism and played a central role in the discovery of the
superfluid phases of $^{3}He$ \cite{betts}. It was generally
believed that this is a unique property of this extremely quantum
system. Recently, we have shown that $YbInCu_{4}$ and related
compounds also present a solid state Pomeranchuk effect
\cite{prb,ssc}. This is best shown in Fig.~\ref{fig1} where at the
first order volume transition temperature, $T_{V}\approx 42$ K, a
local moment phase, indicated by the Curie-Weiss susceptibility,]
coexists with a renormalized Fermi liquid with a nearly
temperature independent susceptibility, such as along the melting
line of $^{3}He$.
An external magnetic field shifts the volume instability to lower
temperatures according to a law of corresponding states
\cite{sarrao,gorkov}. This can be derived from the magnetic
Clausius-Clapeyron equation,
\begin{equation}
\left( \frac{dT}{dH}\right) _{H_{V}}=\frac{-(M_{LM}-M_{FL})_{H_{V}}}{%
(S_{LM}-S_{FL})_{H_{V}}}  \label{CC}
\end{equation}%
At the coexistence line the magnetization of the local moment
phase is much larger than that of the Fermi liquid ($M_{LM} \gg
M_{FL}$). Also, assuming that the entropies in the distinct phases
satisfy the relation $S_{LM} \gg S_{FL}$, we get
\[
\left( \frac{dT}{dH}\right) _{H_{V}}=\frac{-M_{LM}}{S_{LM}}
\]
where $M_{LM}=\chi H=(C/T)H$ and $C$ is the Curie constant.
Substituting in the equation above and integrating yields,
\[
H_{V}^{2}=2\frac{S_{LM}}{C}B-\frac{S_{LM}}{C}T_{V}^{2}
\]
but for $T=0$
\[
H_{V}^{2}(T_{V}=0)=2\frac{S_{LM}}{C}B=H_{0}^{2}
\]
such that,
\begin{equation}
\frac{H_{V}^{2}}{H_{0}^{2}}=1-\frac{T_{V}^{2}}{T_{0}^{2}}.
\label{CL}
\end{equation}%
This is the \emph{circular law of corresponding states}
\cite{sarrao,gorkov} with $T_{0}=\sqrt{C/S_{LM}}H_{0}$. Since
$C=g_{J}^{2}J(J+1)\mu _{B}^{2}/3k_{B}$ and taking $S_{LM}=k_{B}\ln
(2J+1)$ \cite{gorkov}, we find
\[
k_{B}T_{0}=g_{J}\mu _{B}H_{0}\sqrt{\frac{J(J+1)}{3\ln (2J+1)}.}
\]
This yields for the ratio
\begin{equation}
\frac{k_{B}T_{0}}{\mu
_{B}H_{0}}=\frac{g_{J}\sqrt{J(J+1)}}{\sqrt{3\ln
(2J+1)}%
}=\frac{4.5}{\sqrt{3\ln 8}}=\allowbreak 1.\,\allowbreak 80
\label{ratio}
\end{equation}%
in excellent agreement with the experimental result for this ratio
(1.8) \cite {sarrao,gorkov}. We used $J=7/2$. Since $T_{0}=42$ K
\cite{sarrao}, the zero temperature critical field $H_{0}=34.7$ T
in agreement with the scaling result~\cite{sarrao} .

\begin{figure}[th]
\begin{center}
\includegraphics[height=5.6cm]{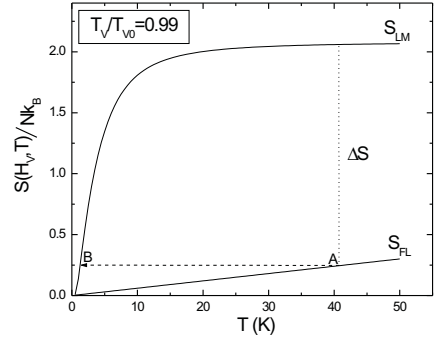}
\end{center}
\caption{Entropy of the local moment (LM) and Fermi liquid phases
(FL) of
$%
YbInCu_{4}$ at a field $H_V$, such
that, $%
T_V(H_V)/T_V(0)=0.99$ \protect\cite{ssc}. The isentropic process
from $A$ to $B$ reduces the temperature of the system.}
\label{fig2}
\end{figure}

\begin{figure}[th]
\begin{center}
\includegraphics[height=4.6cm]{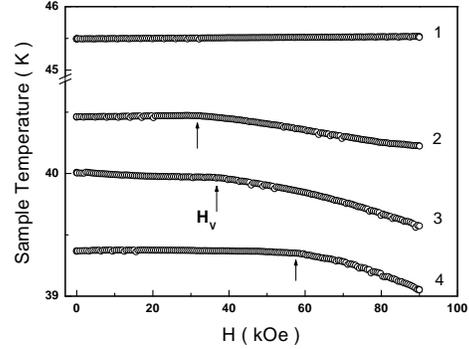}
\end{center}
\caption{ Temperature of the sample under a quasi-adiabatic
application of an external magnetic field. } \label{fig3}
\end{figure}

Fig.~\ref{fig2} shows the entropy of the Fermi liquid and local
moment phases for an external field $H_V=4.9$ T, such that
$T_V(H_V)=0.99T_0$ \cite{ssc}. The entropy of the FL phase is
obtained from the coefficient of the linear term of the specific
heat $\gamma= 50$ mJ/mol K$^2$ \cite{sarrao}. This figure
illustrates how cooling can be produced by adiabatically applying
a magnetic field to the sample in the FL phase and transforming it
to the LM phase. In Fig.~\ref{fig3} we present the results of a
preliminary experiment. The sample temperature under a
quasi-adiabatic application of an external magnetic field,
sweeping from 0 to 9 T, was measured for a single crystal of
$YbInCu{_4} $ (grown as described previously in
Ref.~\cite{sarrao}). In this experiment we used the heat capacity
option setup of a Quantum Design Physical Properties Measurements
System (PPMS). In this setup, the sample is placed on a wired
platform coupled to a temperature sensor and kept under high
vacuum. The sample was zero-field-cooled to a given temperature
near $T_{V}$ and its temperature was measured while the magnetic
field was varied from 0 - 9 T at $\sim$ 100 G/min. The cooling
effect can be clearly observable in the data but it is not of the
expected magnitude. This data provides unambiguous evidence for
the existence of the Pomeranchuk effect in $YbInCu_{4}$.  At the
moment, the magnitude of the effect might be somewhat limited by
the employed experimental conditions. The theoretical expectation
is for a larger effect. Further experiments are in progress to
confirm this.

We have  shown that the solid state Pomeranchuk effect in
$YbInCu_{4}$, can be controlled magnetically and used to produce
cooling.

%

%

\end{document}